\documentclass[lettersize,journal]{IEEEtran}
\usepackage{amsmath,amsfonts}
\usepackage[ruled, linesnumbered]{algorithm2e}
\usepackage{algorithmic}
\usepackage{bm}
\usepackage{array}
\usepackage[caption=false,font=normalsize,labelfont=sf,textfont=sf]{subfig}
\usepackage{textcomp}
\usepackage{stfloats}
\usepackage{url}
\usepackage{verbatim}
\usepackage{graphicx}
\usepackage{cite}
\usepackage{comment}
\usepackage{multirow}
\usepackage{colortbl}
\usepackage[caption=false]{subfig}
\usepackage{booktabs}
\usepackage{hyperref}

\usepackage{balance}

\usepackage{comment}
\usepackage{soul}

\begin{document}

\title{R2C-GAN: Restore-to-Classify Generative Adversarial Networks for Blind X-Ray Restoration and COVID-19 Classification}

\author{Mete Ahishali, Aysen Degerli, Serkan Kiranyaz, Tahir Hamid, Rashid Mazhar, Moncef Gabbouj
\thanks{This work was supported in part by the NSF CBL Program under Project AMaLIA funded by the Business Finland.}
\thanks{Mete Ahishali, Aysen Degerli, and Moncef Gabbouj are with the Faculty of Information Technology and Communication Sciences, Tampere University, Tampere, Finland (email: \textit{name.surname@tuni.fi}).}% <-this % stops a space
\thanks{Serkan Kiranyaz is with the Department of Electrical Engineering, Qatar University, Doha, Qatar (email: \textit{mkiranyaz@qu.edu.qa}).}
\thanks{Tahir Hamid and Rashid Mazhar are with the Hamad Medical Corporation Hospital, Doha, Qatar (email:\textit{ tahirhamid76@yahoo.co.uk} and \textit{rhaq@hamad.qa}).}
}

%\markboth{IEEE Transactions on Neural Networks and Learning Systems,~Vol.~XX, No.~X, XXXX~2023}%

%\IEEEpubid{0000--0000/00\$00.00~\copyright~2021     IEEE}


\maketitle

\begin{abstract}
Restoration of poor quality medical images with a blended set of artifacts plays a vital role in a reliable diagnosis. Existing studies have focused on specific restoration problems such as image deblurring, denoising, and exposure correction where there is usually a strong assumption on the artifact type and severity. As a pioneer study in blind X-ray restoration, we propose a joint model for generic image restoration and classification: \textit{Restore-to-Classify} Generative Adversarial Networks (R2C-GANs). Such a jointly optimized model aims to remove the artifacts while maintaining any disease-related features intact after the restoration. Therefore, this will naturally lead to a higher diagnosis performance thanks to the improved image quality. To accomplish this crucial objective, we define the restoration task as an Image-to-Image translation problem from poor quality having noisy, blurry, or over/under-exposed images to high quality image domain. The proposed R2C-GAN model is able to learn forward and inverse transforms between the two domains using unpaired training samples. Simultaneously, the joint classification preserves the diagnostic-related label during restoration. Moreover, the R2C-GANs are equipped with operational layers/neurons reducing the network depth and further boosting both restoration and classification performances. The proposed joint model is extensively evaluated over the QaTa-COV19 dataset for Coronavirus Disease 2019 (COVID-19) classification. The proposed restoration approach achieves over \boldmath$90\%$ \boldmath$F_1$-Score which is significantly higher than that of any existing deep model. Moreover, in the qualitative analysis, the restoration performance of R2C-GANs is confirmed by a group of medical doctors. We share the software implementation at \href{https://github.com/meteahishali/R2C-GAN}{https://github.com/meteahishali/R2C-GAN}
\end{abstract}


\section{Introduction}
\label{sec:introduction}
\IEEEPARstart{R}{estoration} of poor quality images plays an essential role in many image processing applications. Several different restoration approaches have been proposed revealing the need for a proper restoration technique in different tasks including medical image analysis \cite{chowdhury-enhance}, image classification \cite{rest1}, and remote sensing image classification \cite{rest1, rest2}. It has been shown in these studies that applying image restoration prior to an image processing task improves the overall performance. In general, image restoration encapsulates many different tasks such as image denoising, super-resolution, and general image enhancement, e.g., contrast adjustment or exposure correction.

\begin{figure}[t!]
    \centering
    \includegraphics[width=\linewidth]{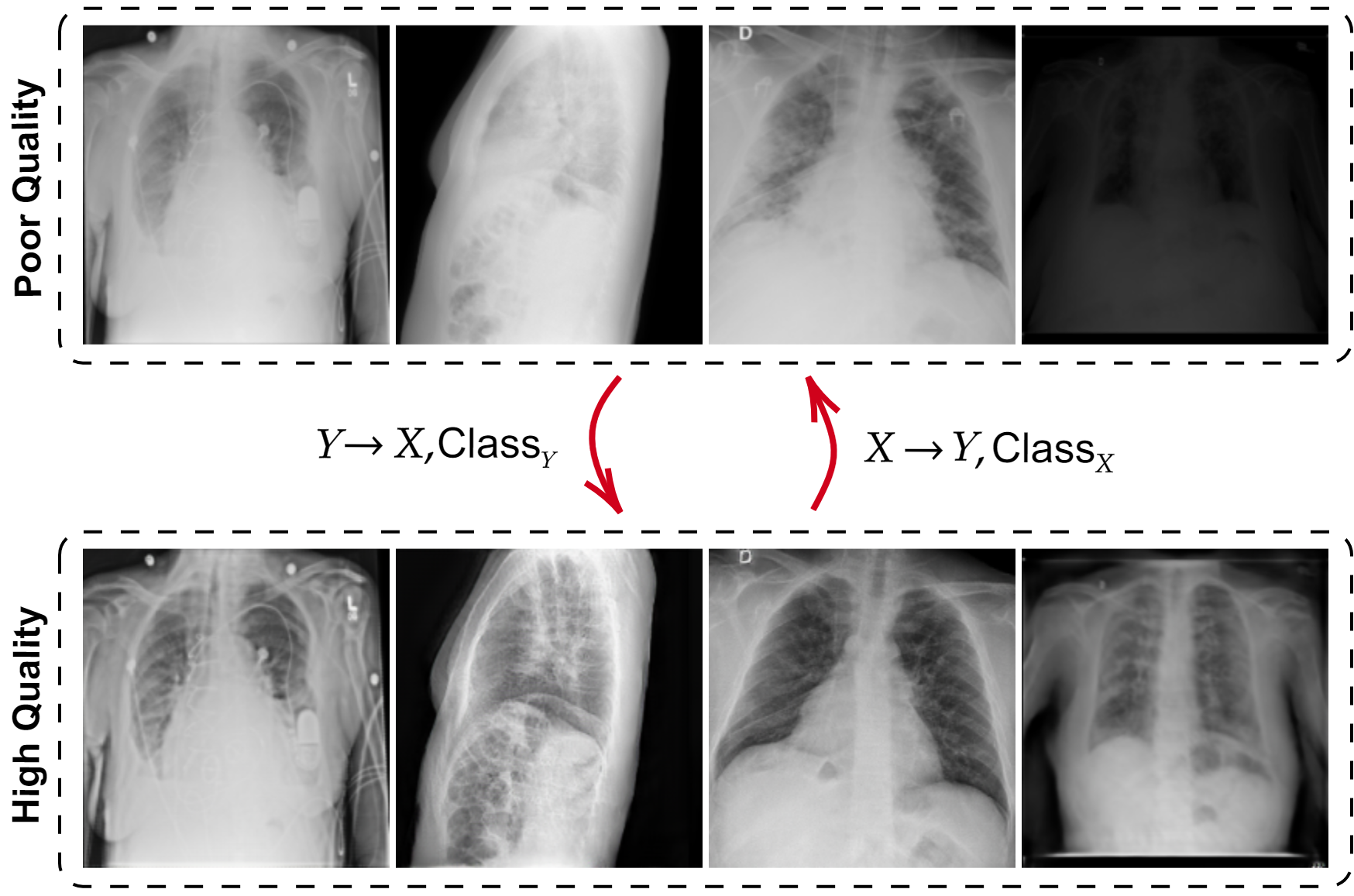}
    \caption{The proposed approach learns forward and inverse transformations between poor and high quality X-ray image domains. During the transitions, disease traces are aimed to be preserved by predicting the class labels.}
    \label{fig:intro}
\end{figure}

\IEEEpubidadjcol

Convolutional Neural Networks (CNNs) have been widely used in image restoration achieving state-of-the-art performances. CNN-based supervised methods require \textit{paired} samples such as Denoising CNNs (DnCNNs) \cite{DnCNNs}, Image-restoration CNNs (IRCNNs) \cite{IRCNN}, denoising autoencoders \cite{denoising-autoencoder}, and Residual Dense Networks (RDNs) \cite{zhang2020residual}. An alternative research direction is called super-resolution where the aim is to find a mapping from low-resolution images to high-resolution images. There have been only a few studies, \cite{ct-superresolution, x-ray-superresolution} focusing solely on medical images, however. A modified U-Net model \cite{ct-superresolution} and a light-weight recursive network \cite{x-ray-superresolution} are proposed for CT and X-ray image super-resolution, respectively. Although such approaches still require paired training samples, it is straightforward to obtain low-resolution images from high quality images by applying down-sampling with a certain kernel function. Finally, it is shown by the study \cite{chowdhury-enhance} that applying different image enhancement techniques, i.e., histogram equalization, gamma correction, balanced contrast enhancement technique (BCET), and contrast limited adaptive histogram equalization (CLAHE) improves the classification performance of the CNNs over chest X-ray images. However, in all such methods, the poor quality samples are artificially created by adding white Gaussian noise, blurring, or corrupting the input image with a known artifact type. Most methods further assume known not only the artifact type but also its properties such as the kernel function for blurring, noise type and variance for denoising. However, in reality, medical images in particular are corrupted by a blend of various artifacts such as noise with varying variances even in the same image, blurring with unknown and possible non-linear kernels, saturation, and lack of contrast. Therefore, the practical usage of such CNN-based solutions is quite limited.

Recently, several methods are proposed where the training data consist of \textit{unpaired} samples. Considering    medical image denoising, few studies \cite{ct-unpaired, x-ray-blind} have proposed blind deep learning based denoising approaches. In \cite{ct-unpaired}, Generative Adversarial Networks (GANs) are used to restore noisy low-dose Computerized Tomography (CT) images. The study achieves blind restoration by introducing the fidelity-embedded GAN (f-GAN) model, where its generator minimizes the weighted combination of two losses: i) the data fidelity loss between low-dose CT (LDCT) image and the restored image and ii) the Kullback-Leibler divergence between the generated distribution and standard-dose CT image data. Another study \cite{x-ray-blind} has proposed to leverage Stein’s Unbiased Risk Estimator (SURE) for blind X-ray denoising Accordingly, let $\mathbf{y}, \mathbf{x} \in \mathbb{R}^{P \times R}$ be noisy and clean images, respectively, such that $\mathbf{y} = \mathbf{x} + \mathbf{n}$ with $\mathbf{n} \sim \mathcal{N}(0, \sigma^2\mathbf{I})$ additive Gaussian noise. The SURE provides an estimation of the mean-squared error $\left \| \mathbf{x} - f(\mathbf{y}) \right \|^2$ for a denoiser network $f$; and hence it eliminates the need for clean ground-truth samples. The proposed approach in \cite{x-ray-blind} is based on training a DnCNN model using SURE loss with unpaired X-ray samples. On the other hand, the SURE estimator has certain drawbacks: i) it is assumed that the noise is additive and Gaussian, and ii) the $\sigma^2$ of the noise is already known. Hence, the model proposed in \cite{x-ray-blind} is only evaluated with artificially degraded clean X-ray samples with known and fixed noise powers (e.g., $\sigma^2$). Therefore, they also suffer from the aforementioned drawbacks of the supervised methods.

In this study, we address this problem as a blind restoration approach thus avoiding any prior assumption over the artifact type, property and severity. Therefore, developing a generic network that can address any restoration problem has the utmost importance, and to the best of the authors' knowledge, such a network has never been proposed in the literature. This study defines the task of image restoration as an Image-to-Image translation problem by learning a mapping function from poor quality to high quality image domain, where the poor quality image samples are noisy, blurry, or over/under-exposed, whereas the high quality images have sharp edges, high-contrast, and crisp with insignificant noise as illustrated in Fig. \ref{fig:intro}. Recently, Cycle-Consistent Generative Adversarial Networks (Cycle-GANs) \cite{cyclegan} have been proposed for the Image-to-Image translation problem. These unsupervised networks learn forward and inverse mappings from a source domain to a target domain without the need for aligned training samples. Although it is still a problem-specific restoration approach, in \cite{cyclegan-dehaze}, a Cycle-GAN based model has achieved improved performance levels in image dehazing. In \cite{cycle-gan-ct}, Cycle‑Deblur GANs have been proposed to learn a mapping from the cone-beam CT (CBCT) domain to the fan-beam CT (FBCT) domain. The study \cite{cycle-gan-ct} has focused mainly on deblurring to restore indistinct CBCT images and performing domain adaption rather than Image-to-Image translation as the selected target and source domains have different acquisition techniques.

Classification of Coronavirus Disease 2019 (COVID-19) using chest X-ray images has been an important task to control the number of active cases. Considering the cost of CT acquisition, automated COVID-19 recognition with X-ray images is more desirable as X-ray devices are highly accessible and fast diagnostic tools. There are several studies \cite{chowdhury-enhance, cov-tnnls, early-cov, seg-cov} focusing on COVID-19 recognition from X-ray images. Accordingly, transfer learning by deep learning models such as EfficientNet \cite{efficientnet}, ResNet \cite{resnet50}, and Inception \cite{inception} trained over ImageNet \cite{imagenet} dataset provides reasonable classification performances. On the other hand, in general, there is a data scarcity problem in medical image analysis, especially for the positive class samples. When the annotated data are limited, it has been observed in \cite{chowdhury-enhance} that enhancing available X-ray samples improves the classification accuracy of different deep learning models.

To accomplish all of the aforementioned blind restoration objectives and to boost the classification performance, we propose \textit{Restore-to-Classify} GANs (R2C-GANs) for blind X-ray image restoration and COVID-19 classification. The R2C-GANs have the unique ability to jointly learn the mapping from poor quality to high/improved quality image domain while performing classification during the transformation as illustrated in Fig. \ref{fig:intro}. The novel training loss and Cycle-GAN model used in the proposed R2C-GANs achieve superior restoration and classification performance levels. It is worth noting that in blind X-ray image restoration, preserving the patient-specific information within the lung area is challenging since unpaired samples are used in the training. This pilot study shows that combining the classification and restoration tasks in a single network enables the preservation of the infection/disease information in the domain transitions. As a result of this mutual relationship, while the restoration performance is improved due to the introduced classification task in the network, the COVID-19 classification performance is also improved thanks to the applied restoration.

\begin{figure*}
         \centering
         \subfloat[\label{fig:framework_a}]{\includegraphics[width=.47\linewidth]{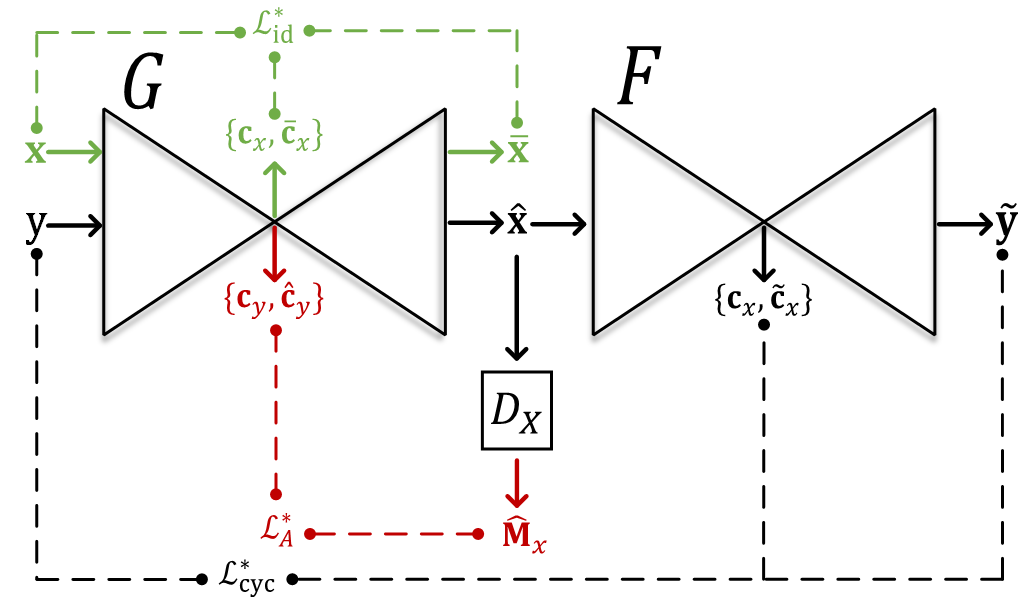}}
    \hfill
         \centering
         \subfloat[\label{fig:framework_b}]{\includegraphics[width=.47\linewidth]{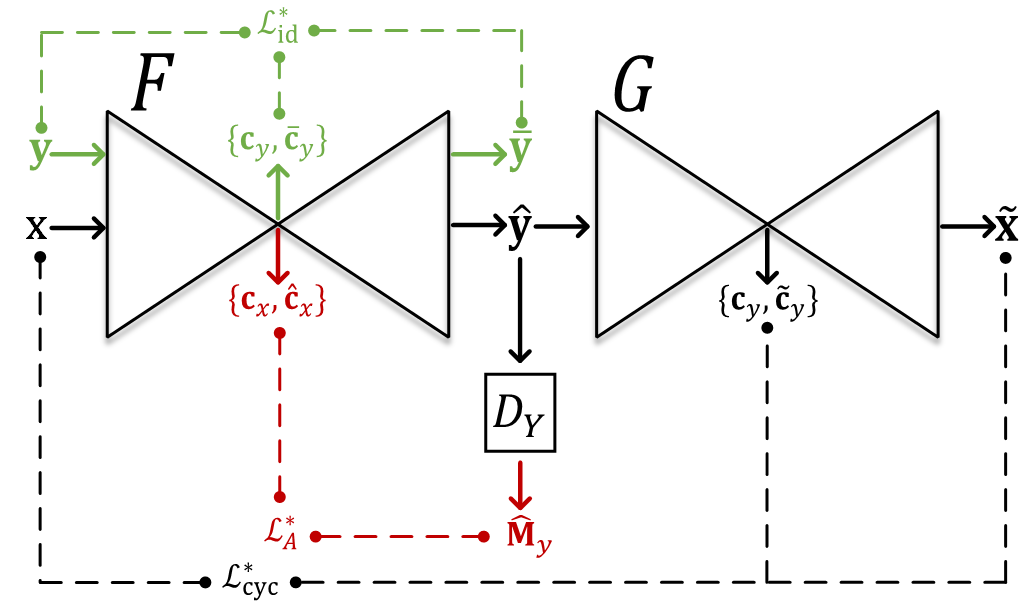}}
     \centering
     \caption{The R2C-GAN framework is divided into two parts for the illustration purposes. There are two generator networks providing $G: Y \rightarrow X, C_Y$ and $F: X \rightarrow Y, C_X$ where $Y$ and $X$ are poor and high quality image domains, respectively, with $\mathbf{y} \in Y$, $\mathbf{x} \in X$. Their class labels are $\mathbf{c}_{y} \in C_Y$ and $\mathbf{c}_{x} \in C_X$ where $C_Y, C_X \in \mathbb{R}^{N_C}$ with $N_C$ number of classes. Two discriminators learn the following mappings: $D_X: X \rightarrow M_X$ and $D_Y: Y \rightarrow M_Y$ where $M_X, M_Y \in \mathbb{R}^{d_m \times d_n}$ are the predicted masks.}
     \label{fig:framework}
\end{figure*}

In the literature, several studies \cite{j3, j1, j2} have proposed task-oriented regression frameworks. In \cite{j3}, a denoiser autoencoder is jointly trained with a classifier for lung-nodule classification using LDCT images. However, unlike the proposed R2C-GANs, their framework includes a two-stage/sequential mapping by attaching the classifier to the output of the denoiser autoencoder. Similar multi-staged sequential approaches are proposed for denoising LDCT images \cite{j1} and Magnetic Resonance Imaging (MRI) reconstruction \cite{j2}. The main difference is that \cite{j1} and \cite{j2} use paired-samples for training. Therefore, the corrupted images are artificially synthesized from high quality images. However, most actual artifacts cannot be generated by this approach especially for X-rays. Additionally, in \cite{j1}, input images are not real images, instead, they are synthesized from high-quality CT images. In \cite{j2}, the framework is proposed for MRI acquisition, where they try to reconstruct MR images from undersampled $k$-space data, i.e., compressive sensing. Most important of all, both CT and MRI are significantly better (high quality and resolution) data acquisition methodologies compared to X-ray imaging where severe artifacts do usually occur. Finally, both studies \cite{j1, j2} focus on segmentation whereas our focus in this study is the classification of X-ray images.

To further boost both restoration and classification performances, the proposed R2C-GAN model is designed with self-organized operational layers \cite{onn1}. The Self-Organized Operational Neural Networks (Self-ONNs) are proposed in \cite{onn1} with the "generative" neuron model where non-linear kernel transformation functions are approximated using Taylor-series expansion and the corresponding polynomial coefficients become the trainable parameters. Thus, the operational layers have a superior ability to learn highly non-linear mappings contrary to the traditional convolutional layers. The Self-ONNs have been evaluated in different tasks \cite{onn1, onn3, onn2, onn4} achieving a significant superiority with an elegant computational efficiency. The proposed R2C-GAN model with the operational layers can therefore be shallow to prevent over-fitting in the presence of scarce data. Even with such compact architectures, the R2C-GANs have a superior learning capability compared to the competing deep CNNs with a large number of convolutional layers.

Overall, the contributions of this study can be summarized as follows:
\begin{itemize}
    \item Contrary to the prior studies focusing on the restoration of a  single artifact type in a supervised manner, the proposed restoration approach can blindly learn the mapping from poor quality to high quality image domain. Hence, the proposed approach can perform restoration regardless of the blend of artifacts and their severities.
    \item The proposed network model, called R2C-GANs, can perform a joint blind restoration and classification. Therefore, the proposed approach is a "goal-oriented" restoration method exploiting the relation between image restoration and classification.
    \item R2C-GANs are equipped with operational layers as an alternative to the convolutional layers. Hence, they can learn highly non-linear mappings between the poor quality and high quality image domains. Additionally, the compact R2C-GAN architectures achieve computational efficiency thanks to the proposed compact architecture.
    \item An extensive set of evaluations over the QaTa-COV19 X-ray image dataset \cite{seg-cov} has demonstrated that the proposed R2C-GANs achieve the \textit{state-of-the-art} classification performance outperforming recent deep networks including EfficientNet-B4, EfficientNet-B5, ResNet-50, and Inception-v3.
    \item Qualitative evaluations by a group of Medical Doctors (MDs) have shown that the restored X-rays are preferable most of the time compared to the original X-rays; and the proposed approach significantly improves the lung segmentation performance of X-ray images when the lung segmentation model is applied over the restored images using R2C-GANs.
\end{itemize}

The rest of the paper is organized as follows: the proposed methodology is presented in Section \ref{sec:method}. The experimental setup and an extensive set of three-fold evaluations (quantitative, qualitative, and medical) are covered in Section \ref{sec:results}. Finally, Section \ref{sec:conc} concludes the paper and suggests topics for the future research.

\section{Proposed Methodology}
\label{sec:method}

In this section, we will briefly introduce the self-organized operational layers with generative neurons. The R2C-GANs with their implementation for blind restoration and classification will be detailed next.

\subsection{Self-Organized Operational Layers}

In Self-ONNs \cite{onn1,onn3,onn2,onn4}, the non-linear kernel transformation function for each neuron is approximated using $Q^{\text{th}}$ order Taylor series expansion. Hence, the neurons or the kernel elements are \textit{self-organized} during the training procedure as each kernel element's transformation function is different and self-learned since they are formed based on the converged coefficient values.

Taylor series of a function $f$ at a single point $y$ near the origin is given in the form of an infinite sum of its derivatives,
\begin{equation}
    f(y) = \sum_{n=0}^{\infty}\frac{f^{(n)}(0)}{n!}(y)^n,
\end{equation}
while the $Q^{\text{th}}$ order approximation (i.e., Taylor polynomial) can be expressed as follows:
\begin{equation}
    f(y)^{(Q)} = \sum_{q=0}^{Q}\frac{f^{(q)}(0)}{q!}(y)^q.
\end{equation}
Let $w_q = \frac{f^{(q)}(0)}{q!}$ be the $q^{th}$ coefficient of the $Q^{th}$ order polynomial. Accordingly, a \textit{generative} neuron in operational layers performs the following transformation:
\begin{equation}
    g(y, \mathbf{w}) = w_0 + yw_1 + y^2w_2 + \dots + y^Qw_Q,
\end{equation}
where $\mathbf{w} \in \mathbb{R}^Q$ is the trainable parameter, omitting $w_0$ since there is a common bias term applied after the transformation.

In 2D operational layers, the trainable parameter set of the $k^{\text{th}}$ filter is expressed as $\mathbf{W}^{(k)} = [\mathbf{w}_1^{(k)}, \mathbf{w}_2^{(k)}, \dots, \mathbf{w}_Q^{(k)}] \in \mathbb{R}^{m \times n \times Q}$, where $\mathbf{w}_q^{(k)} \in \mathbb{R}^{m \times n}$ is the $k^{\text{th}}$ operational filter with a size of $m \times n$ for the $q^{\text{th}}$ coefficient. For instance, assuming a single-channel input $\mathbf{y}^{(k)} \in \mathbb{R}^{P \times R}$ for the $k^{\text{th}}$ filter, a pixel of the corresponding output is computed as follows,
\begin{equation}
    x_{p, r}^{(k)} = \sum_{q=0}^Q \sum_{i=0}^{m-1} \sum_{j=0}^{n-1} w_{q, i, j}^{(k)}\left(y_{p + i, r + j}^{(k)} \right)^q.
\end{equation}
Note that since the summations are commutative, the above expression can be expressed by the summation of the consecutive convolutions and biases, $b_q^{(k)}$,
\begin{equation}
    \label{eq:operation}
    \mathbf{x}^{(k)} = \sigma \left( \sum_{q=1}^Q \left(\mathbf{y} ^{(k)} \right)^q * \mathbf{w}_q^{(k)} + b_q^{(k)} \right),
\end{equation}
where $\sigma(.)$ is the activation function and $*$ is the convolution operation. Therefore, in the proposed layer, the $k^{\text{th}}$ filter will have the following learnable parameters $\mathbf{\Theta}_k = \{\mathbf{W}^{(k)} \in \mathbb{R}^{m \times n \times Q}, \mathbf{b}^{(k)} \in \mathbb{R}^{Q}\}$.

\subsection{Restore-to-Classify GANs (R2C-GANs)}
The proposed \textit{restore-to-classify} approach with the R2C-GANs consists of two generator and two discriminator Self-ONNs as depicted in Fig. \ref{fig:framework}. The proposed approach has a novel improved generator mapping ability compared to the conventional Cycle-GANs so that during the forward and inverse mappings, R2C-GANs can estimate the class labels of the source domain images. Moreover, with the operational layers embedded into the R2C-GAN model, the performance is further improved.

\begin{figure}[t!]
    \centering
    \includegraphics[width=0.5\textwidth]{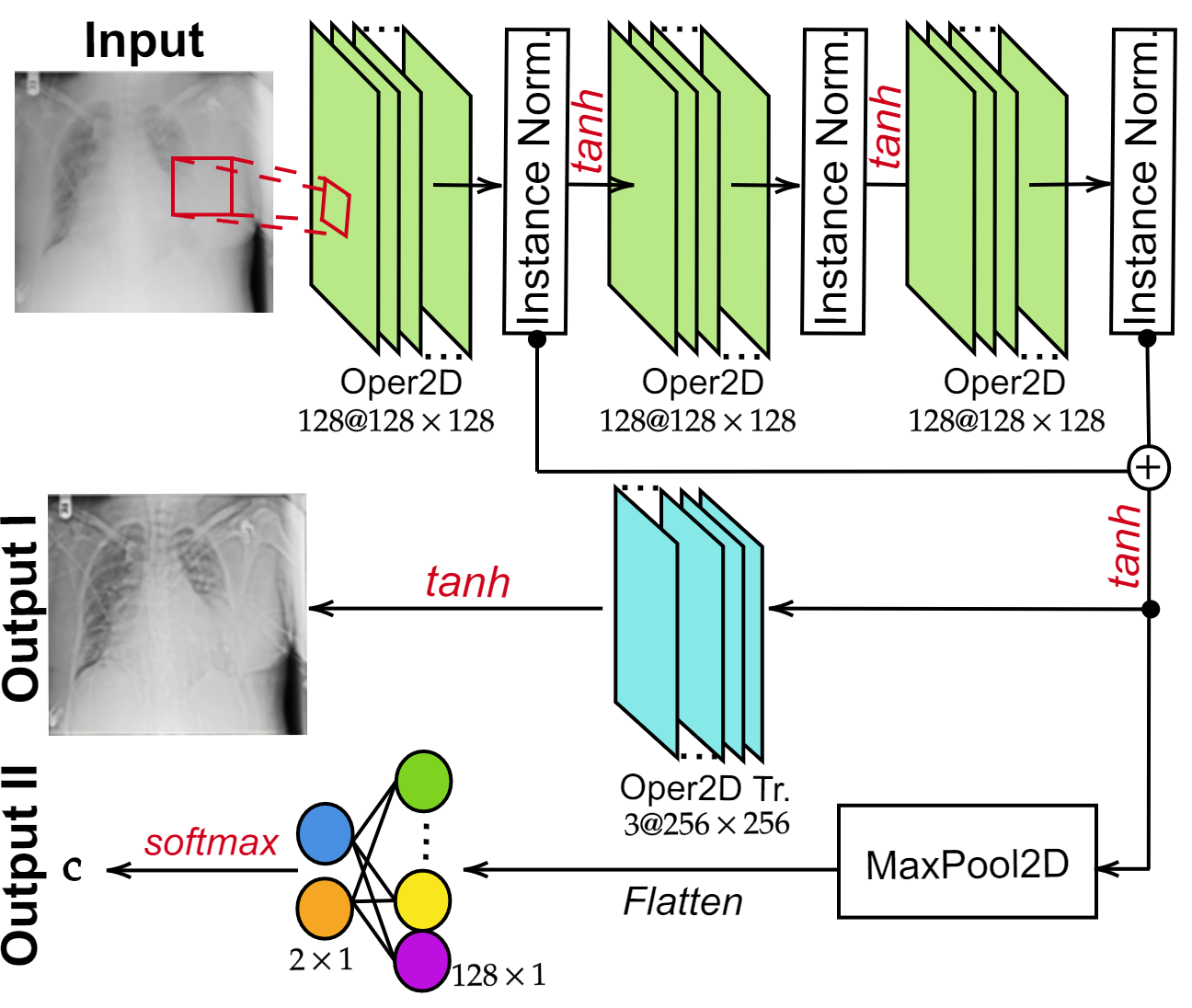}
    \caption{The R2C-GANs have the presented compact and novel structure for their generator networks $G$ and $F$. The proposed generator model applies two tasks in a single inference: Image-to-Image translation and classification.}
    \label{fig:generator}
\end{figure}

\begin{figure}[t!]
    \centering
    \includegraphics[width=0.5\textwidth]{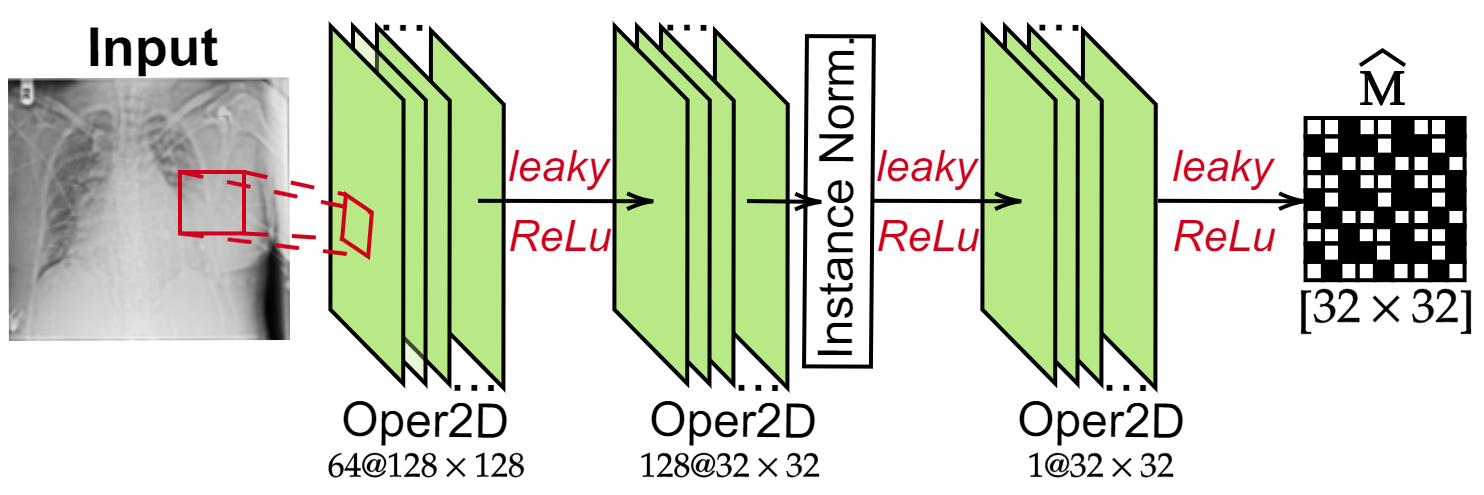}
    \caption{The proposed compact and novel network structure is illustrated for the discriminator networks $D_X$ and $D_Y$ in R2C-GANs.}
    \label{fig:discriminator}
\end{figure}

Having the sets of poor quality samples, $\{\mathbf{y}_i\}_{i=1}^{N_p}$ (e.g., corrupted, noisy, blurry, saturated, low-contrast or low resolution images), and high quality samples, $\{\mathbf{x}_j\}_{j=1}^{N_h}$ (e.g., crisp, sharp, high-contrast and without any saturation), we define two image domains such that $\mathbf{y}_i \in Y$ and $\mathbf{x}_j \in X$. Correspondingly, their class labels (one hot encoded) form the following domains: $\mathbf{c}_{y_i} \in C_Y$ and $\mathbf{c}_{x_j} \in C_X$ with $C_Y, C_X \in \mathbb{R}^{N_C}$, where $N_C$ is the number of classes. In R2C-GANs, two generator networks learn the forward and inverse mappings between the image domains and provide the predicted class labels during the domain transition process such that $G: Y \rightarrow X, C_Y$ and $F: X \rightarrow Y, C_X$. Both generators have the same network configuration as illustrated in Fig. \ref{fig:generator}. The novel design for the proposed generators includes only four operational layers and one dense layer. There are also two discriminator networks, $D_Y$ and $D_X$ the architecture of which is shown in Fig. \ref{fig:discriminator}. They are trained to determine whether the given input image is generated (synthetic) or is a real sample from the target domain. Accordingly, in one cycle of the proposed R2C-GAN model starting with forward mapping, given a corrupted poor quality image $\mathbf{y}$, the restored image and its predicted label are first produced: $G(\mathbf{y}) = \left\{ \hat{\mathbf{x}}, \hat{\mathbf{c}}_y \right\}$. Next, the corresponding discriminator is fed by the restored image: $D_X(\hat{\mathbf{x}}) = \hat{\mathbf{M}}_x$ where $\hat{\mathbf{M}}_x \in \mathbb{R}^{d_m \times d_n}$ is the computed mask denoting pixel-wise prediction. The dimension of the estimated mask depends on the number of downsampling operations. Finally, one-cycle is completed by reconstructing the input image using the restored image with the inverse mapping: $F(\hat{\mathbf{x}}) = \left\{ \tilde{\mathbf{y}}, \hat{\mathbf{c}}_x \right\}$. Similarly, the second cycle is computed by choosing $X$ as the source domain and starting with the generator $F$ as given in Fig. \ref{fig:framework_b}. Then, the discriminator $D_Y$ is used and finally, the reconstruction is applied by the generator $G$.

The objective function of the proposed R2C-GAN approach consists of four different losses including adversarial, cycle-consistency, identity, and classification losses; $\mathcal{L}_A$, $\mathcal{L}_{\text{cyc}}$, $\mathcal{L}_{\text{id}}$, $\mathcal{L}_{\text{class}}$, respectively:
\begin{equation}
\label{eq:objective}
    \begin{split}
    \mathcal{L}_{\text{T}}(G, F, D_Y, D_X, & \mathbf{y}, \mathbf{x}) = \mathcal{L}_A(G, D_X, \mathbf{y}, \mathbf{x}) \\
    & + \mathcal{L}_A(F, D_Y, \mathbf{x}, \mathbf{y}) + \lambda \mathcal{L}_{\text{cyc}}(G, F, \mathbf{x}, \mathbf{y}) \\
    & + \beta \mathcal{L}_{\text{id}}(G, F, \mathbf{x}, \mathbf{y}) + \gamma \mathcal{L}_{\text{class}}(G, F, \mathbf{x}, \mathbf{y}).
    \end{split}
\end{equation}

During the training process of GANs, the generator and discriminator networks are trained in an adversary way. For instance, the weights of the generator $G$ and $D_X$ are updated such that $\min_G \mathcal{L}_A(G, D_X, \mathbf{y}, \mathbf{x})$ and $\max_{D_X} \mathcal{L}_A(G, D_X, \mathbf{y}, \mathbf{x})$. Thus, the forward mapping generator tries to transform poor quality images to high quality images in such a way that the transformed images are very similar to the real high quality images from $X$ domain so that the discriminator cannot distinguish them.

\subsubsection{Training Procedure of the Generator Networks}

Let us first define the modified adversarial, cycle, and identity losses in the proposed R2C-GAN model. Next, we will form the complete loss function in a more general form as in \eqref{eq:objective}. Given unpaired samples from the training data, $\mathbf{y} \in Y$ and $\mathbf{x} \in X$, the generators perform the Image-to-Image translation and classification:  $G(\mathbf{y}) = \left\{ \hat{\mathbf{x}}, \hat{\mathbf{c}}_y \right\}$ and $F(\mathbf{x}) = \left\{ \hat{\mathbf{y}}, \hat{\mathbf{c}}_x \right\}$. Then, the following losses are computed:
\begin{equation}
\label{eq:adv-cost1}
    \mathcal{L}_A^*(D_X, \hat{\mathbf{x}}, \mathbf{c}_y, \hat{\mathbf{c}}_y) = \left \| D_X\left( \hat{\mathbf{x}} \right) - \mathbf{1} \right \|_2^2 - \lambda_a \sum_{i=1}^{N_C} c_{y,i} \log(\hat{c}_{y,i}),
\end{equation}
\begin{equation}
\label{eq:adv-cost2}
    \mathcal{L}_A^*(D_Y, \hat{\mathbf{y}}, \mathbf{c}_x, \hat{\mathbf{c}}_x) = \left \| D_Y\left(\hat{\mathbf{y}} \right) - \mathbf{1} \right \|_2^2 - \lambda_a \sum_{i=1}^{N_C} c_{x,i} \log(\hat{c}_{x,i}),
\end{equation}
where $\mathbf{c}_y, \mathbf{c}_x \in \mathbb{R}^{N_c}$ are the true class vectors of $\mathbf{y}$ and $\mathbf{x}$, respectively. In the next iteration, we compute the following maps to finish two one-cycles from both ends: $G(\hat{\mathbf{y}}) = \left\{ \tilde{\mathbf{x}}, \tilde{\mathbf{c}}_x \right\}$ and $F(\hat{\mathbf{x}}) = \left\{ \tilde{\mathbf{y}}, \tilde{\mathbf{c}}_y \right\}$. Using the modified cycle-consistency loss, we penalize the reconstruction and classification errors of two inputs for both ends: $\mathbf{y} \rightarrow \hat{\mathbf{x}} \rightarrow \tilde{\mathbf{y}}, \tilde{\mathbf{c}}_y$ and $\mathbf{x} \rightarrow \hat{\mathbf{y}} \rightarrow \tilde{\mathbf{x}}, \tilde{\mathbf{c}}_x$:
\begin{equation}
\label{eq:cyc-cost}
    \begin{split}
    \mathcal{L}_{\text{cyc}}^*(\mathbf{y}, \tilde{\mathbf{y}}, \mathbf{c}_y, \tilde{\mathbf{c}}_y, \mathbf{x}, \tilde{\mathbf{x}}, \mathbf{c}_x, \tilde{\mathbf{c}}_x) = \left \| \tilde{\mathbf{y}} - \mathbf{y} \right \|_1 + \left \| \tilde{\mathbf{x}} - \mathbf{x} \right \|_1 \\
    - \lambda_c \sum_{i=1}^{N_C} c_{y,i} \log(\tilde{c}_{y,i}) - \lambda_c \sum_{i=1}^{N_C} c_{x,i} \log(\tilde{c}_{x,i}).
    \end{split}
\end{equation}

Finally, in the third step, we feed both generators with their target domain image samples. Then, the generators are forced to produce the same input image at the output such that $G(\mathbf{x}) = \left\{ \bm \bar{\mathbf{x}}, \bm \bar{\mathbf{c}}_x \right\}$ and $F(\mathbf{y}) = \left\{ \bm \bar{\mathbf{y}}, \bm \bar{\mathbf{c}}_y \right\}$. This is accomplished by using the modified identity loss where we include the classification error as well,
\begin{equation}
\label{eq:id-cost}
    \begin{split}
    \mathcal{L}_{\text{id}}^*(\mathbf{x}, \bm \bar{\mathbf{x}}, \mathbf{c}_x, \bm \bar{\mathbf{c}}_x, \mathbf{y}, \bm \bar{\mathbf{y}}, \mathbf{c}_y, \bm \bar{\mathbf{c}}_y) = \left \| \bm \bar{\mathbf{x}} - \mathbf{x} \right \|_1 + \left \| \bm \bar{\mathbf{y}} - \mathbf{y} \right \|_1 \\
    - \lambda_i \sum_{i=1}^{N_C} c_{x,i} \log\bm (\bar{c}_{x,i}) - \lambda_i \sum_{i=1}^{N_C} c_{y,i} \log\bm (\bar{c}_{y,i}).
    \end{split}
\end{equation}
It is essential in an automated image restoration task that if the input image belongs to the high quality set, the restoration framework should keep the input image unchanged after the transformation is applied. In R2C-GANs, the additional classification term included in the identity loss further ensures preserving \textit{class-specific} features in the input images.

\begin{algorithm}[t!]
\caption{Training Procedure of the proposed approach with R2C-GANs.}
\label{alg:training}
\textbf{Input:} Training samples $\mathbf{y}_t, \mathbf{x}_t$, $\lambda$, $\beta$, $\gamma$, \textit{maxIter}, and learning parameters. \\
\textbf{Output:} Trained generator and discriminator weights: $\mathbf{\Theta}_G, \mathbf{\Theta}_F, \mathbf{\Theta}_{D_X}$ and $\mathbf{\Theta}_{D_Y}$. \\
Initialize the trainable parameters, $\mathbf{\Theta}_G, \mathbf{\Theta}_F, \mathbf{\Theta}_{D_X}, \mathbf{\Theta}_{D_Y}$; \\
Apply normalization over $\mathbf{y}_t$ and $\mathbf{x}_t$; \\

\While{ $t < $ maxIter}{
Sample $\mathbf{y}, \mathbf{x}$ from the train set;\\

\tcc{Generator training:}
Obtain $G_{(t)}(\mathbf{y}) \rightarrow \left\{ \hat{\mathbf{x}}, \hat{\mathbf{c}}_y \right\}$ and $F_{(t)}(\mathbf{x}) \rightarrow \left\{ \hat{\mathbf{y}}, \hat{\mathbf{c}}_x \right\}$; \\
Obtain $F_{(t)}(\hat{\mathbf{x}}) \rightarrow \left\{ \tilde{\mathbf{y}}, \tilde{\mathbf{c}}_y \right\}$ and $G_{(t)}(\hat{\mathbf{y}}) \rightarrow \left\{ \tilde{\mathbf{x}}, \tilde{\mathbf{c}}_x \right\}$; \\
Obtain $G_{(t)}(\mathbf{x}) \rightarrow \left\{ \bm \bar{\mathbf{x}}, \bm \bar{\mathbf{c}}_x \right\}$ and $F_{(t)}(\mathbf{y}) \rightarrow \left\{ \bm \bar{\mathbf{y}}, \bm \bar{\mathbf{c}}_y \right\}$; \\

Compute: $L_1 = \mathcal{L}_A^*(D_{X_t}, \hat{\mathbf{x}}, \mathbf{c}_y, \hat{\mathbf{c}}_y) + \mathcal{L}_A^*(D_{Y_t}, \hat{\mathbf{y}}, \mathbf{c}_x, \hat{\mathbf{c}}_x)$ using \eqref{eq:adv-cost1} and \eqref{eq:adv-cost2}; \\
Compute: $L_2 = \mathcal{L}_{\text{cyc}}^*(\mathbf{y}, \tilde{\mathbf{y}}, \mathbf{c}_y, \tilde{\mathbf{c}}_y, \mathbf{x}, \tilde{\mathbf{x}}, \mathbf{c}_x, \tilde{\mathbf{c}}_x)$ $\eqref{eq:cyc-cost}$; \\
Compute: $L_3 = \mathcal{L}_{\text{id}}^*(\mathbf{x}, \bm \bar{\mathbf{x}}, \mathbf{c}_x, \bm \bar{\mathbf{c}}_x, \mathbf{y}, \bm \bar{\mathbf{y}}, \mathbf{c}_y, \bm \bar{\mathbf{c}}_y)$ $\eqref{eq:id-cost}$; \\
Calculate the updated weights $\mathbf{\Theta}_G(t + 1)$ and $\mathbf{\Theta}_F(t + 1)$ minimizing $L_G = L_1 + \lambda L_2 + \beta L_3$ using ADAM optimizer; \\

\tcc{Discriminator training:}
Obtain new mappings using updated generators: $G_{(t + 1)}(\mathbf{y}) \rightarrow \left\{ \hat{\mathbf{x}}, \_ \right\}$ and $F_{(t + 1)}(\mathbf{x}) \rightarrow \left\{ \hat{\mathbf{y}}, \_ \right\}$; \\
Obtain $D_{Y_{(t)}}(\mathbf{y}) \rightarrow \hat{\mathbf{M}}_{y, r}$ and $D_{Y_{(t)}}(\hat{\mathbf{y}}) \rightarrow \hat{\mathbf{M}}_{y, f}$; \\
Obtain $D_{X_{(t)}}(\mathbf{x}) \rightarrow \hat{\mathbf{M}}_{x, r}$ and $D_{X_{(t)}}(\hat{\mathbf{x}}) \rightarrow \hat{\mathbf{M}}_{x, f}$; \\
Compute: $L_D = \left\| \hat{\mathbf{M}}_{y, r} - \mathbf{1} \right\|_2^2 + \left\| \hat{\mathbf{M}}_{y, f} \right\|_2^2 + \left\| \hat{\mathbf{M}}_{x, r} - \mathbf{1} \right\|_2^2 + \left\| \hat{\mathbf{M}}_{x, f} \right\|_2^2$; \\
Calculate the updated model weights $\mathbf{\Theta}_{D_X}(t + 1)$ and $\mathbf{\Theta}_{D_Y}(t + 1)$ that minimizes $L_D$ using ADAM \cite{adam} optimizer; \\
}
\end{algorithm}

Overall, the training procedure of the R2C-GANs is presented in Algorithm \ref{alg:training}, where we indicate the generator and discriminator training sections separately. In the generator training, the minimized total loss $L_G$ is the summation of the losses computed in \eqref{eq:adv-cost1}, \eqref{eq:adv-cost2}, \eqref{eq:cyc-cost}, and \eqref{eq:id-cost}. By a proper selection of the regularization parameters, the total sum can be expressed in the form of \eqref{eq:objective}, i.e., setting $\lambda_a = \gamma$, $\lambda_c = \gamma/\lambda$, and $\lambda_i = \gamma/\beta$, then, $L_G$ is expressed as,
\begin{equation}
\label{eq:total_gen_cost}
    \begin{split}
    L_G = \left \| D_X\left( \hat{\mathbf{x}} \right) - \mathbf{1} \right \|_2^2 + \left \| D_Y\left(\hat{\mathbf{y}} \right) - \mathbf{1} \right \|_2^2 \\
    + \lambda \left( \left \| \tilde{\mathbf{y}} - \mathbf{y} \right \|_1 + \left \| \tilde{\mathbf{x}} - \mathbf{x} \right \|_1 \right) \\
    + \beta \left( \left \| \bm \bar{\mathbf{x}} - \mathbf{x} \right \|_1 + \left \| \bm \bar{\mathbf{y}} - \mathbf{y} \right \|_1 \right) \\
    + \gamma \biggl[ \sum_{i=1}^{N_C} - \biggl( c_{y,i} \log(\hat{c}_{y,i}) + c_{x,i} \log(\hat{c}_{x,i}) \\
    + c_{y,i} \log(\tilde{c}_{y,i}) + c_{x,i} \log(\tilde{c}_{x,i}) \\
    + c_{x,i} \log(\bm \bar{c}_{x,i}) + c_{y,i} \log(\bm \bar{c}_{y,i}) \biggl) \biggl].
    \end{split}
\end{equation}

\subsubsection{Training Procedure of the Discriminator Networks}

The training of the discriminators in R2C-GANs is straightforward and similar to GANs. Given $\mathbf{y}$ and $\mathbf{x}$, the adversarial losses are computed as follows:
\begin{equation}
\label{eq:dis1}
    \mathcal{L}_A(G, D_X, \mathbf{y}, \mathbf{x}) = \left\| D_X(\mathbf{x}) - \mathbf{1} \right\|_2^2 + \left\| D_X(G(\mathbf{y}))\right\|_2^2
\end{equation}
\begin{equation}
\label{eq:dis2}
    \mathcal{L}_A(F, D_Y, \mathbf{x}, \mathbf{y}) = \left\| D_Y(\mathbf{y}) - \mathbf{1} \right\|_2^2 + \left\| D_Y(F(\mathbf{x}))\right\|_2^2
\end{equation}
Then, the discriminator weights are computed minimizing $L_D = \mathcal{L}_A(G, D_X, \mathbf{y}, \mathbf{x}) + \mathcal{L}_A(F, D_Y, \mathbf{x}, \mathbf{y})$.

As presented in Algorithm \ref{alg:training}, the generators and discriminators are trained in an adversarial manner. In one training iteration, the generator models are updated to trick their corresponding discriminators by forcing them to produce the mask of $\mathbf{1}$ when they are fed by the transformed images. This behaviour is due to the minimization of \eqref{eq:adv-cost1} and \eqref{eq:adv-cost2}. Then, the transformed images are computed again using the updated generator models. The discriminators are updated to produce the output mask of $\mathbf{1}$ for the real images, i.e., minimizing \eqref{eq:dis1} and \eqref{eq:dis2}.

\section{Experimental Evaluation}
\label{sec:results}

In this section, we first detail the experimental setup used in the evaluation including the benchmark dataset, hyperparameter values, and computational and software environments. Next, the quantitative and qualitative results are presented followed by the computational complexity analysis.

\subsection{Experimental Setup}

The QaTa-COV19 dataset \cite{seg-cov,osegnet} consists of chest X-ray images with $10,323$ COVID-19 and nearly $130,000$ control group samples including healthy subjects and $14$ different thoracic diseases. The progress of COVID-19 varies among the positive class samples, for example, early-stage where the infection trace is not visible or very limited, mild-stage, and severe-stage samples. These samples are mainly collected from the following sources: BIMCV-COVID19+ dataset \cite{bimcv}, Italian Society of Medical and Interventional Radiology (SIRM) \cite{sirm}, and Hannover Medical School and Institute for Diagnostic and Interventional Radiology \cite{hannover}. The majority of the control group samples are from the ChestX-ray14 dataset \cite{chestx-ray8}. For the complete list of the sources and more information about the dataset, the readers are referred to the dataset repository\footnote{The QaTa-COV19 benchmark dataset is available at the following repository: https://www.kaggle.com/datasets/aysendegerli/qatacov19-dataset.} and the studies \cite{seg-cov} and \cite{osegnet}.

\begin{figure}[t!]
         \centering
         \subfloat[\label{fig:dataset_a}]{\includegraphics[width=0.49\linewidth]{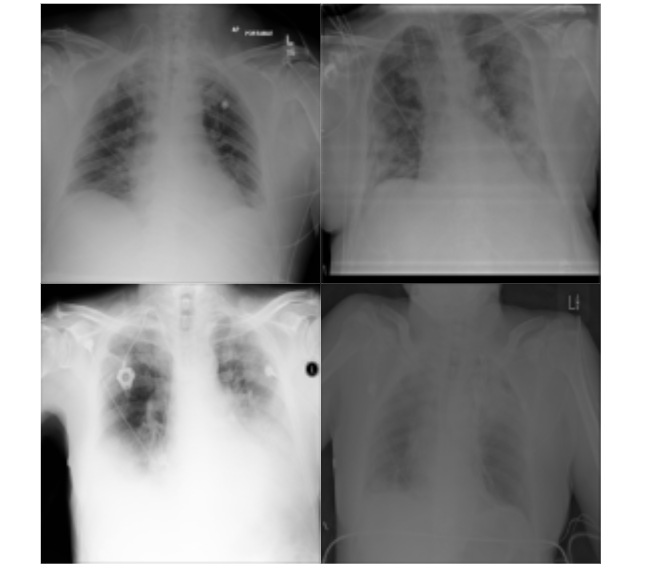}}
         \centering
         \subfloat[\label{fig:dataset_b}]{\includegraphics[width=0.49\linewidth]{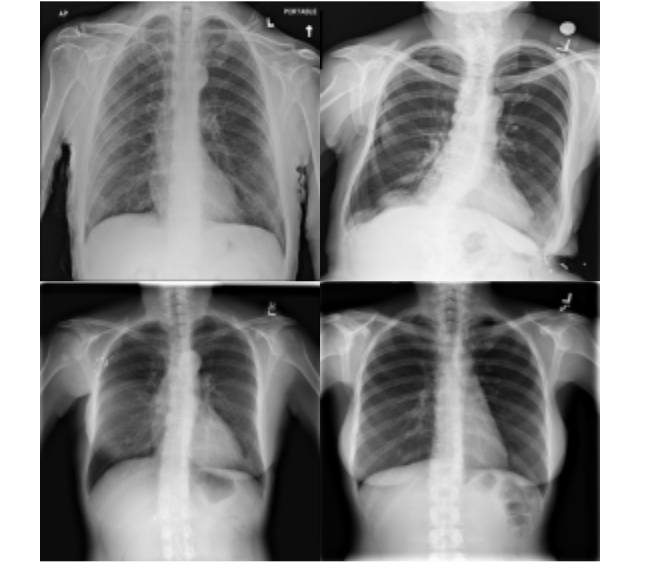}}
     \centering
     \caption{Poor quality (a) and high quality (b) image samples from the QaTa-COV19 dataset.}
     \label{fig:dataset}
\end{figure}

For the training of the proposed approach, we have investigated both positive and control group samples of the QaTa-COV19 dataset. By considering their overall qualities, we have manually annotated a set of samples and grouped them in two categories: poor quality samples $\{\mathbf{y}_i\}_{i=1}^{N_p}$ and high quality samples $\{\mathbf{x}_j\}_{j=1}^{N_h}$. Accordingly, we have labeled $N_p = 2460$ poor quality and $N_h = 2094$ high quality samples. Samples from the dataset are provided in Fig. \ref{fig:dataset}. Basically, assigning the training samples defines the transformation domains that is going to be learned by the R2C-GAN; therefore, it has a significant effect on the restoration performance. To avoid ambiguity in image domains, we have mainly excluded the mid-quality images in the formation of the training set. In the test phase, the restoration and classification performance evaluations have been performed by randomly selecting $19,247$ images from the QaTa-COV19 dataset. Overall, Table \ref{tab:dataset} presents the number of samples in the selected subset. As the test samples are randomly selected, the image quality varies in the test set, so it may contain poor, medium, and high quality images.

\begin{table}[t!]
\centering
\caption{Number of poor and high quality training samples and test samples per class.}
\label{tab:dataset}
\resizebox{\linewidth}{!}{
\begin{tabular}{|c|c|c|c|c|c|}
\hline
\rowcolor[gray]{.92}Samples & \begin{tabular}[c]{@{}c@{}}Image \\ Quality\end{tabular} & COVID-19 & Control Group & Total \\ \hline \hline
\multirow{2}{*}{Training} & Poor & $851$ & $1069$ & $2460$ \\ 
&  High & $505$ & $1589$ &$2094$ \\ \hline \hline
\rowcolor[gray]{.97}Testing & All & $3247$ & $16 000$ & $19 247$ \\ \hline
\end{tabular}}
\end{table}

The experimental evaluations have been performed using Tensorflow library \cite{tensorflow} on Python. The gradient tracing functionality of the library is utilized in the R2C-GAN implementation. We publicly share the software implementation and the restoration dataset annotations\footnote{The software implementation of the proposed R2C-GANs is available at https://github.com/meteahishali/R2C-GAN.}. The computational complexity analysis has been performed on a computing node having Intel ® Xeon 4214 CPU with 64 GB memory and NVidia ® Tesla V100 GPU. The ADAM optimizer \cite{adam} has been used in the training procedure with the following empirically set learning parameters: the decay rates of the momentum updates are $\beta_1 = 0.5$, $\beta_2 = 0.999$ and learning rate $\alpha = 2 \times 10^{-4}$. The R2C-GAN model is trained for 2000 epochs and similar to \cite{cyclegan}, we linearly decay the learning rate starting from the $100^{\text{th}}$ epoch such that $\alpha \rightarrow 0$ as the training proceeds. In \eqref{eq:total_gen_cost}, we set $\lambda = 10$ and $\beta = 5$, i.e., the proposed values in the initial Cycle-GANs \cite{cyclegan}. On the other hand, the introduced classification trade-off parameter is empirically set as $\gamma = 0.1$; and correspondingly, $\lambda_a = \gamma = 0.1$, $\lambda_c = \gamma/\lambda = 0.01$, and $\lambda_i = \gamma/\beta = 0.02$.

\begin{table*}[t!]
\caption{COVID-19 classification results using proposed R2C-GANs and competing deep networks. The R2C-GAN (CNN) has two configurations, one is more compact (*).}
\label{tab:class_results}
\centering
\resizebox{.8\linewidth}{!}{
\begin{tabular}{cccccccc}
\toprule
 & \multicolumn{1}{c}{\textbf{Method}} & \multicolumn{1}{c}{\textbf{Accuracy}} & \multicolumn{1}{c}{\textbf{Sensitivity}} & \multicolumn{1}{c}{\textbf{Specificity}} & \multicolumn{1}{c}{\textbf{Precision}} & \multicolumn{1}{c}{\boldmath$F_1$\textbf{-Score}} & \multicolumn{1}{c}{\boldmath$F_2$\textbf{-Score}} \\ \midrule

\multicolumn{1}{c}{\multirow{5}{*}{\rotatebox[origin=c]{90}{\textit{\textbf{Deep}}}}}\hspace{-0.3cm} & EfficientNet-B4 & $0.6013$ & $0.9923$ & $0.5219$ & $0.2964$ & $0.4564$ & $0.6752$ \\

\multicolumn{1}{c}{} & EfficientNet-B5 & $0.9116$ & $0.9593$ & $0.9019$ & $0.6649$ & $0.7854$ & $0.8813$ \\

\multicolumn{1}{c}{} & ResNet-50 & $0.6236$ & \boldmath$0.9945$ & $0.5484$ & $0.3088$ & $0.4713$ & $0.6887$ \\

\multicolumn{1}{c}{} & Inception-v3 & $0.9483$ & $0.8950$ & $0.9591$ & $0.8161$ & $0.8537$ & $0.8780$ \\

\multicolumn{1}{c}{} & R2C-GAN (CNN) & \boldmath$0.9744$ & $0.9095$ & \boldmath$0.9876$ & \boldmath$0.9369$ & \boldmath$0.9230$ & \boldmath$0.9148$ \\ \midrule

\multicolumn{1}{c}{\multirow{4}{*}{\rotatebox[origin=c]{90}{\textit{\textbf{Compact}}}}}\hspace{-0.3cm} & R2C-GAN (CNN*) & $0.9529$ & $0.8476$ & $0.9743$ & $0.8701$ & $0.8587$ & $0.8520$ \\

\multicolumn{1}{c}{} & R2C-GAN ($Q = 1$) & $0.9370$ & $0.8223$ & $0.9603$ & $0.8079$ & $0.8150$ & $0.8194$ \\

\multicolumn{1}{c}{} & R2C-GAN ($Q = 3$) & $0.9586$ & $0.8867$ & $0.9733$ & $0.8706$ & $0.8785$ & $0.8834$ \\

\multicolumn{1}{c}{} & R2C-GAN ($Q = 5$) & \boldmath$0.9676$ & \boldmath$0.8968$ & \boldmath$0.9820$ & \boldmath$0.9100$ & \boldmath$0.9034$ & \boldmath$0.8994$ \\ \bottomrule
\end{tabular}}
\end{table*}

\subsection{Results}

\subsubsection{Quantitative Analysis}

In COVID-19 classification performance evaluation, accuracy is computed by the ratio of the number of correctly classified samples to the total number of samples,
\begin{equation}
    \text{Accuracy} = (\text{TP} + \text{TN}) / (\text{TP} + \text{TN} + \text{FP} + \text{FN}),
\end{equation}
where TP and TN are the number of true positive and control group samples, respectively, and FP and FN are the number of misclassified control group and positive class samples, respectively. Next, specificity is defined as the ratio of TN to the number of control group samples:
\begin{equation}
    \text{Specificity} = \text{TN} / (\text{TN} + \text{FP}),
\end{equation}
whereas the ratio of TP to the number of positive samples and to the number of predicted positive samples define the sensitivity and precision, respectively:
\begin{equation}
    \text{Sensitivity} = \text{TP} / (\text{TP} + \text{FN}),
\end{equation}
\begin{equation}
    \text{Precision} = \text{TP} / (\text{TP} + \text{FP}).
\end{equation}
Finally, $F_1$ and $F_2$ scores are computed as follows with $\beta = 1$ and $\beta = 2$, respectively:
\begin{equation}
    F_\beta = (1 + \beta ^ 2) \frac{\text{Precision} \times \text{Sensitivity}}{\beta ^ 2 \times \text{Precision} + \text{Sensitivity}}.
\end{equation}
Consequently, $F_2$ score has more significance on reduced FN than FP.

COVID-19 classification results are presented in Table \ref{tab:class_results}. In the results, the R2C-GAN (CNN) and R2C-GAN (CNN*) models use convolutional layers in the proposed restore-to-classify approach. The top-5 models have deep configurations whereas the rest of the methods on the bottom are compact. The last deep CNN inherits the initial design proposed by \cite{cyclegan}, which is significantly deeper than the other R2C-GANs, e.g., it has $12$ times more parameters than the R2C-GAN (with $Q = 3$). The classification branch providing Output II in Fig. \ref{fig:generator} is connected to the output of the fifth ResNet block in the generator networks of the R2C-GAN (CNN) model. On the other hand, R2C-GAN (CNN*) has a similar configuration to R2C-GAN ($Q = 1$) except that the activation functions of the generators are set to ReLu. Overall, it is observed in Table \ref{tab:class_results} that the proposed restore-to-classify approach outperforms all competing deep networks in COVID-19 classification with a significant gap. Among the competitors, only Inception-v3 can achieve an $F_1$-score above $85\%$. Comparing different R2C-GAN models, compact networks with operational layers obtain comparable classification performance with the deep R2C-GAN (CNN) model. Especially, the R2C-GAN model with $Q = 5$ achieves around $90\%$ $F_1$ and $F_2$-scores.

\subsubsection{Qualitative Restoration Evaluations}

%ROW 1 COVID-19 (d), ROW 2 COVID-19 (d), ROW 3 COVID-19 (d), ROW 4 no finding (d), ROW 5 infiltration\&mass (d)
\begin{figure*}[t!]
    \centering
    \includegraphics[width=0.93\linewidth]{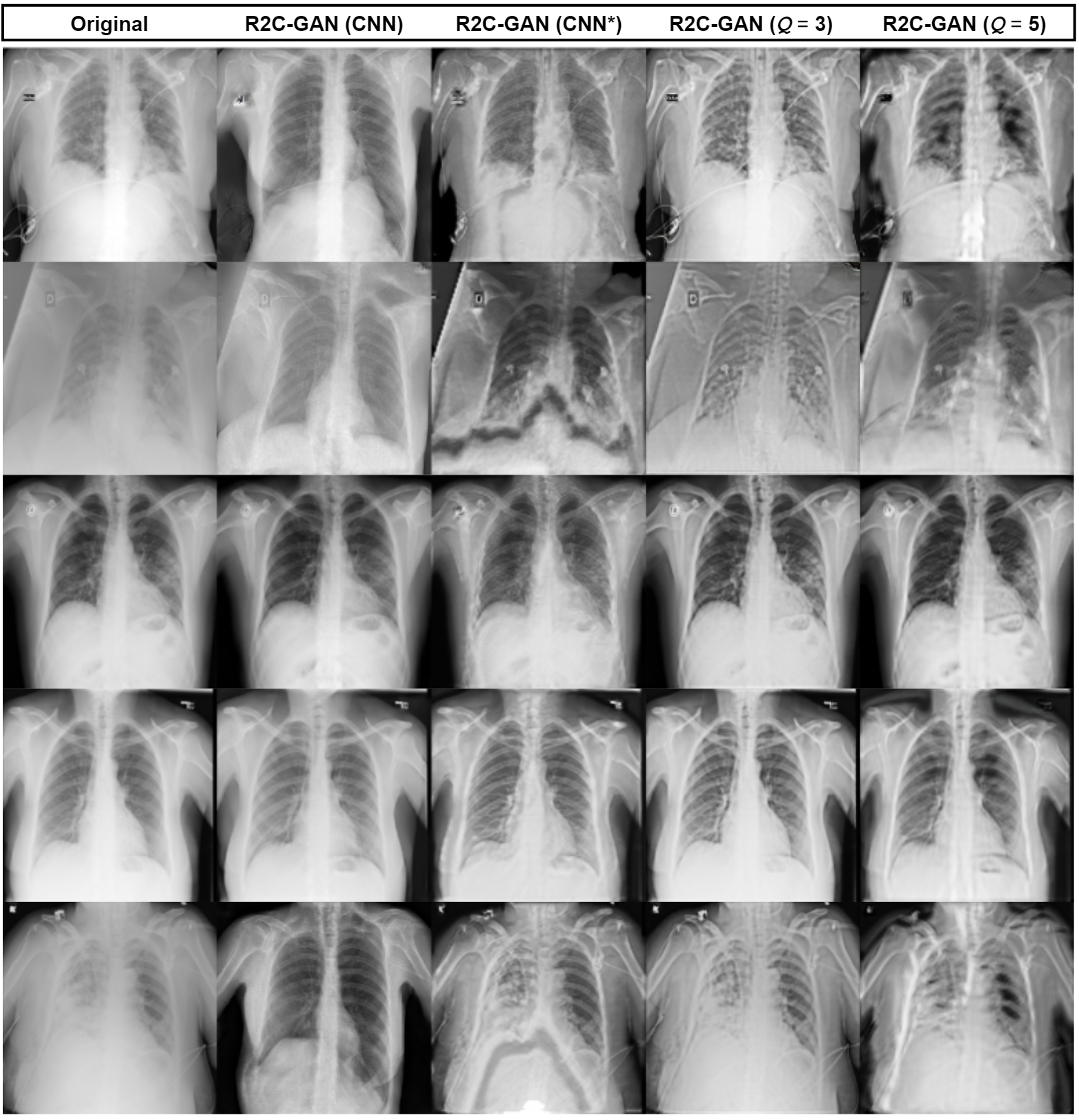}
    \caption{Original samples with the corresponding restorations of different R2C-GAN models are shown in each column.}
    \vspace{0.2cm}
    \label{fig:res_results}
\end{figure*}

The qualitative evaluation is performed over $2000$ randomly selected samples from the test set. The original and its corresponding four different restored X-ray images obtained from R2C-GAN (CNN), R2C-GAN (CNN*), R2C-GAN ($Q = 3$), and R2C-GAN ($Q = 5$) models are used for the evaluation. Accordingly, we split the samples between four MDs who are experts in the field of COVID-19 diagnosis from chest X-images and they are asked which X-ray image (the original or four different restored versions) they would prefer for the diagnosis considering the provided disease labels. In Fig. \ref{fig:res_results}, one row represents one query that is presented to the MD for voting. Note that for a given row, the selection is performed blindly so that the MDs do not have the information about the column labels. Then, the selection ratio is computed for each restored version including the original images as presented in Table \ref{tab:res_results}. It is observed that using operational layers in R2C-GANs with $Q = 3$ achieves the highest overall restoration performance. Due to the subjective evaluation, several MDs have preferred the original images for the diagnosis, while MD III and IV have voted more for the restored images by R2C-GAN ($Q=3$) model than for the original images with selection ratios larger than $55\%$. We have observed that during inference, iterative restoration of the images improves the performance of the compact R2C-GAN models, whereas the deep R2C-GAN model introduces more blur in the iterative restoration. Therefore, the presented restored images to the MDs are obtained by consequently applying the generators three times, i.e., for a given image sample $\mathbf{y}$, the generator $G$ is applied three times: $G \circ G \circ G(\mathbf{y})$ while the generator of deep R2C-GAN (CNN) is applied only once. In Fig. \ref{fig:res_results}, it can be observed that satisfactory restoration results are obtained using R2C-GAN (CNN) and R2C-GAN ($Q = 3$). While the R2C-GAN (CNN*) model produces certain artifacts, R2C-GANs ($Q = 3$) restores without any artifacts and the restored images are shaper with significantly more details. Considering the fifth row in Fig. \ref{fig:res_results}, R2C-GAN (CNN) removes the disease information on lung areas and makes the region more blackish. Finally, it is shown that when the input image has already a high overall quality as in the third row, the proposed approach tends to keep the high quality after the transformation applied. We have also observed that the proposed restoration approach occasionally removes a part of a text or sign on the X-rays. It is important to note that the restoration aims to present an alternative but mostly superior visualization of the original X-rays for better diagnostics and analysis. If such an occasional restoration artifact is encountered, it can, therefore, be easily compensated by the MDs by analyzing the original X-ray.

\begin{table}[ht!]
\vspace{0.2cm}
\caption{Quality Restoration Evaluation: The Selection Ratio (Number of Selections / Total Selections) of Each MD.}
\vspace{0.1cm}
\label{tab:res_results}
\centering
\begin{tabular}{@{}cccccc@{}}
\toprule
\multirow{2}{*}{\textbf{MD}} & \multirow{2}{*}{Original} & R2C-GAN & R2C-GAN & R2C-GAN & R2C-GAN \\
  &          & (CNN) & (CNN*)  & ($Q=3$)   & ($Q=5$) \\ \midrule
I & $0.4323$ & $0.2085$ & $0.0502$ & $0.2329$ & $0.0761$ \\
II & $0.5162$ & $0.2130$ & $0.0090$ & $0.2491$ & $0.0126$ \\
III & $0.210$ & $0.160$ & $0.0640$ & $0.5580$ & $0.0080$ \\
IV & $0.0433$ & $0.3258$ & $0.0557$ & $0.5629$ & $0.0124$ \\ \midrule
Overall & $0.3169$ & $0.2245$ & $0.0442$ & \boldmath$0.3839$ & $0.0305$ \\ \bottomrule
\end{tabular}
\end{table}

\begin{table}[t!]
\vspace{0.2cm}
\caption{Segmentation Evaluation: The Selection Ratio (Number of Selection / Total Selections) of Each MD.}
\vspace{0.1cm}
\label{tab:seg_results}
\centering
\begin{tabular}{@{}cccccc@{}}
\toprule
\multirow{2}{*}{\textbf{MD}} & \multirow{2}{*}{Original} & R2C-GAN & R2C-GAN & R2C-GAN & R2C-GAN \\
 & & (CNN) & (CNN*)  & ($Q=3$)   & ($Q=5$)   \\ \midrule
I & $0.3837$ & $0.3531$ & $0.0323$ & $0.1732$ & $0.0577$ \\
II & $0.2271$ & $0.2310$ & $0.1564$ & $0.2167$ & $0.1687$ \\
III & $0.3691$ & $0.4787$ & $0.0716$ & $0.1186$ & $0.0313$\\
IV & $0.7085$ & $0.0583$ & $0.0135$ & $0.2152$ & $0.0045$ \\ \midrule
Overall & \boldmath$0.3497$ & $0.2660$ & $0.0986$ & $0.1935$ & $0.1025$ \\ \bottomrule
\end{tabular}
\end{table}

\newpage
\begin{figure}[b!]
    \centering
    \includegraphics[width=\linewidth]{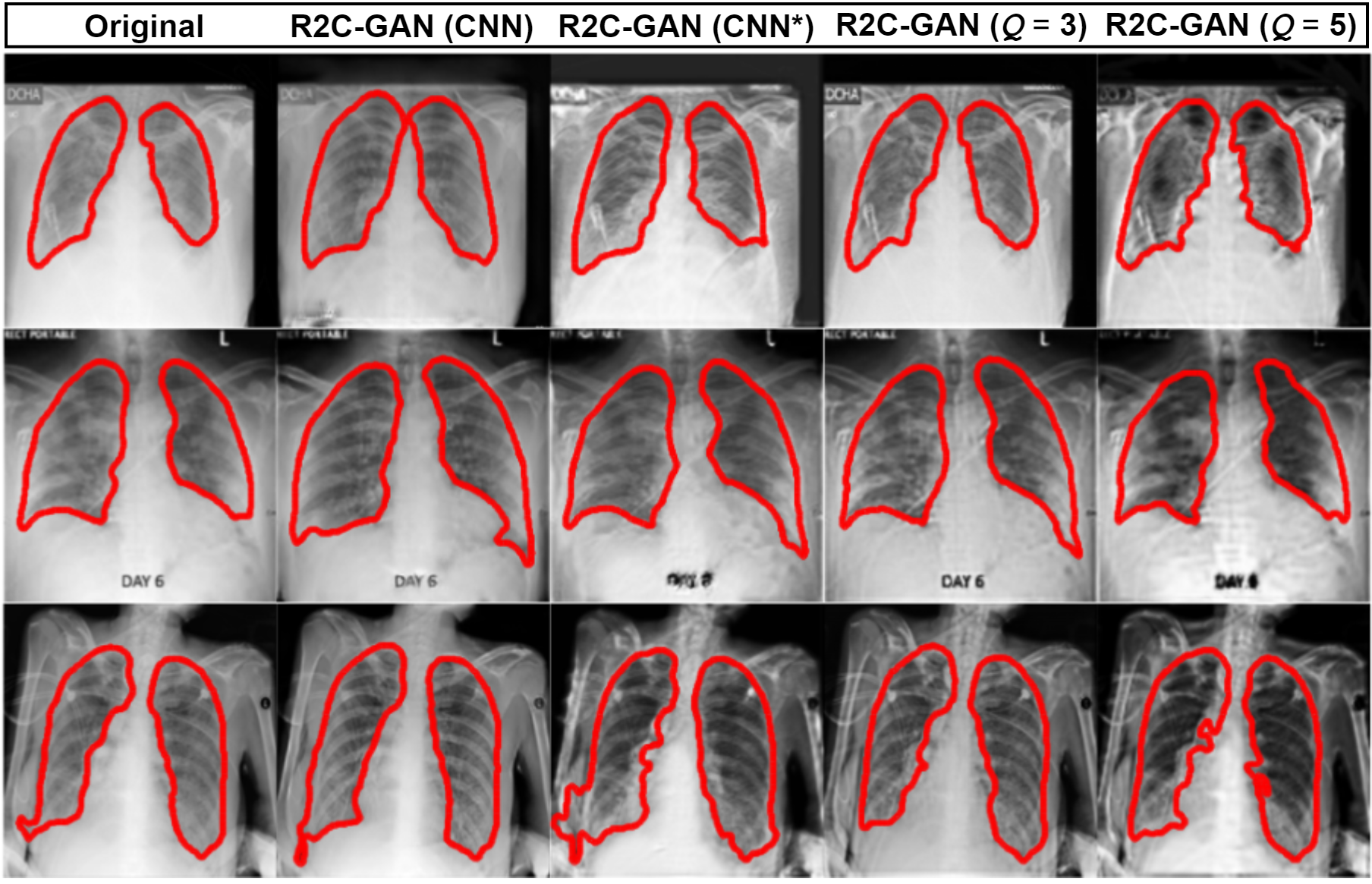}
    \caption{Segmented lung regions are illustrated when the U-Net model is applied over the original and corresponding restored samples.}
    \label{fig:seg_results}
\end{figure}

%\textcolor{red}{ROW 1 COVID-19 (d,d), ROW 2 COVID-19 (d,d), ROW 3 mass\&nodule (d,d)}
    
Additionally, the MDs are asked to evaluate the accuracy of segmentation masks obtained by a pre-trained U-Net model \cite{unet} over the original and restored images. We use DenseNet-121 \cite{DenseNet} as the backbone of U-Net model and its weights are initialized with ImageNet weights. Then, it is fine-tuned using Montgomery \cite{Montgomery} and JSRT datasets \cite{jsrt, jsrt-mask} consisting of a total number of $385$ X-ray images with their corresponding lung segmentation masks. The model is trained for $15$ epochs with a learning rate of $\alpha = 10^{-4}$ and ADAM optimizer. Finally, the lung areas of $2000$ images from the initial test set of R2C-GANs have been computed using the fine-tuned segmentation model. As before, for a given row in Fig. \ref{fig:seg_results}, the MDs selections are requested blindly for $2000$ queries. Correspondingly, the selection ratios are presented in Table \ref{tab:seg_results}. The total number of selections for the segmented lung areas using original images is only 34.97\%, while the remaining selections has favored the restored images. The selections of MD III for the restored images are larger than $10\%$ than the original segmented images. It is also observed that with the proposed operational layers ($Q = 3$), R2C-GAN greatly improves the performance of the compact R2C-GAN model with convolutional layers. Interestingly, the R2C-GAN ($Q = 3$) model can achieve comparable segmentation improvements with the deep R2C-GAN (CNN) model in the selections of the MD II. As illustrated in Fig. \ref{fig:seg_results}, the segmented lung regions are greatly improved by the proposed approach compared to the original images, especially by the R2C-GAN (CNN) and R2C-GAN ($Q = 3$) models.

\vspace{0.3cm}
\subsubsection{Computational Complexity Analysis}
\label{sec:computational}

The number of trainable parameters and average elapsed times per test sample are presented in Table \ref{tab:complexity} for all methods. Accordingly, the competing deep networks require more memory, their time complexity are comparable with the R2C-GAN models. It has been observed that R2C-GAN ($Q=3$) model has given comparable restoration performances with R2C-GAN (CNN), whilst it has significantly reduced computational complexity than the R2C-GAN (CNN) model in terms of both memory and time.

\begin{table}[h!]
\centering
\vspace{0.2cm}
\caption{The number of trainable parameters and averaged elapsed times/per sample are given for the competing deep networks and the proposed R2C-GANs (Classification and Restoration + Classification).}
\vspace{0.1cm}
\label{tab:complexity}
\resizebox{.95\linewidth}{!}{
\begin{tabular}{c|c|cc|}
\cline{2-4}
 & \multicolumn{1}{c|}{\textbf{Model}} & \multicolumn{1}{c}{\textbf{Parameters}} & \multicolumn{1}{c|}{\textbf{Time (ms)}}  \\ \hline

\multicolumn{1}{|c|}{\multirow{9}{*}{{\rotatebox[origin=c]{90}{\textit{Classification}}}}} & EfficientNet-B4 & $17.552$ M & $2.164$ \\ 

\multicolumn{1}{|c|}{} &\cellcolor[gray]{.98} EfficientNet-B5 & \cellcolor[gray]{.98}$28.345$ M & \cellcolor[gray]{.98}$2.477$ \\ 

\multicolumn{1}{|c|}{} & ResNet-50 & $23.539$ M & $1.572$ \\ 

\multicolumn{1}{|c|}{} & \cellcolor[gray]{.98} Inception-v3 & \cellcolor[gray]{.98}$21.772$ M & \cellcolor[gray]{.98}$1.165$ \\ 

\multicolumn{1}{|c|}{} & R2C-GAN (CNN) & $6.283$ M & $3.490$ \\ 

\multicolumn{1}{|c|}{} & \cellcolor[gray]{.98} R2C-GAN (CNN*) & \cellcolor[gray]{.98}$0.299$ M & \cellcolor[gray]{.98}$1.663$ \\ 

\multicolumn{1}{|c|}{} & R2C-GAN ($Q = 1$) & $0.299$ M & $1.594$ \\ 

\multicolumn{1}{|c|}{} & \cellcolor[gray]{.98} R2C-GAN ($Q = 3$) & \cellcolor[gray]{.98}$0.896$ M & \cellcolor[gray]{.98}$2.172$ \\

\multicolumn{1}{|c|}{} & R2C-GAN ($Q = 5$) & $1.493$ M & $3.339$ \\ \hline \hline

\multicolumn{1}{|c|}{\multirow{5}{*}{\rotatebox[origin=c]{90}{\textit{Res. + Class.}}}} & \cellcolor[gray]{.98} R2C-GAN (CNN) & \cellcolor[gray]{.98}$11.384$ M & \cellcolor[gray]{.98}$8.142$ \\ 

\multicolumn{1}{|c|}{} & R2C-GAN (CNN*) & $0.303$ M & $3.0$ \\ 

\multicolumn{1}{|c|}{} & \cellcolor[gray]{.98} R2C-GAN ($Q = 1$) & \cellcolor[gray]{.98}$0.303$ M & \cellcolor[gray]{.98}$3.113$ \\ 

\multicolumn{1}{|c|}{} & R2C-GAN ($Q = 3$) & $0.907$ M & $3.739$ \\ 

\multicolumn{1}{|c|}{} & \cellcolor[gray]{.98} R2C-GAN ($Q = 5$) & \cellcolor[gray]{.98}$1.510$ M & \cellcolor[gray]{.98}$4.928$ \\ \hline

\end{tabular}}
\end{table}

\subsection{Ablation Study: Restoration without Classification}

\begin{figure}[h!]
    \centering
    \includegraphics[width=0.95\linewidth]{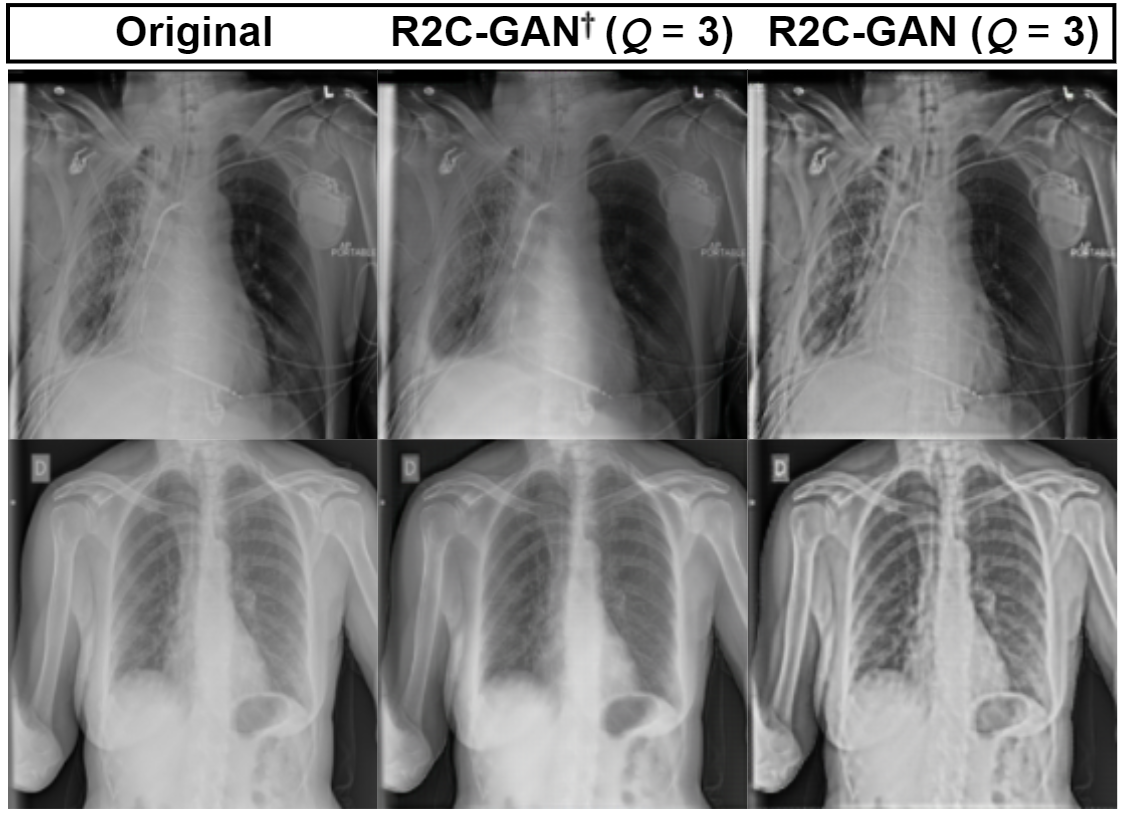}
     \caption{Restoration of original X-rays (left) by the proposed $\text{R2C-GAN}^\dagger \text{(}Q=3\text{)}$ without (middle) and with (right) classification.}
    \label{fig:abl_results}
\end{figure}

It is important to show that when the network is trained only for the restoration task without the proposed restore-to-classify framework, the performance is significantly decreased. In this manner, we set $\gamma = 0$ in the loss function \eqref{eq:total_gen_cost} of the generators and present the restored images in Fig. \ref{fig:abl_results}. It can be observed that the classification plays an important role in the restoration performance since the transformed images show no improvement in quality when the classification part of the R2C-GAN is disabled.

\section{Conclusion}
\label{sec:conc}

Image restoration is an essential task since their poor quality can highly degrade the performance of many analysis tasks such as segmentation and classification. In particular, the restoration of medical images is crucial considering the data scarcity and critical effect of various artifacts on the diagnosis. In this study, we propose a network model for joint restoration and classification of X-ray images, i.e., in a single inference, the network restores the input image and performs classification. More importantly, R2C-GANs can perform a blind restoration without requiring any prior assumption over the artifact types and their severities. We have shown that the proposed approach can preserve the disease information after the transformation with the proposed restore-to-classify learning approach. Overall, this is a pioneer study where a blind and goal-oriented image restoration approach is proposed and applied in the absence of paired training data. As a joint regression and classification network model, this study has demonstrated that classification improves the regression (restoration) performance, and \textit{vice versa}. Furthermore, we significantly reduce the computational complexity by utilizing operational layers in the proposed compact R2C-GAN models. The experimental evaluations have shown that the proposed approach significantly outperforms the recent deep networks: EfficientNet-B4, EfficientNet-B5, ResNet-50, and Inception-v5 in COVID-19 classification. In the qualitative analysis, the MDs have approved the restoration performance by visually comparing the restored images and examining the lung regions extracted by the pre-trained U-Net model. Thanks to its generic design, R2C-GANs can easily be applied to other corrupted biomedical signals. This will be the topic of our future research.

\section*{Acknowledgment}
We would like to thank Muhammad Muslim (MD), Consultant Pulmonology/Thoracic Surgery, Hamad Medical Corporation, Al Wakrah, Qatar, and Samman Rose (MD), Fellow Internal Medicine, Hamad General Hospital, Doha, Qatar, for their contribution in the qualitative evaluation.
 
\bibliographystyle{IEEEtran}
\bibliography{IEEEtran}
\balance

\vfill

\end{document}